  \documentclass[structabstract]{aa}  
\usepackage{graphicx}
\usepackage{txfonts}
\usepackage{natbib}
\bibpunct{(}{)}{;}{a}{}{,}
\newcommand\linea{Fe~{\sc i}~$\lambda$6301.5\AA}
\newcommand\lineb{\ion{Fe}{i}~$\lambda$6302.5\,\AA}
\newcommand\paperiii{SA05}

\newcommand\hinode{{\em Hinode}}
\begin{document}
\title{On the origin of reverse polarity patches found by {\em Hinode} 
	in sunspot penumbrae}
   \author{J.~S\'anchez~Almeida\inst{1,2} \and 
	K.~Ichimoto\inst{3}}
    \institute{
    	Instituto de Astrof\'\i sica de Canarias, 
        E-38205 La Laguna, Tenerife, Spain\\
	\email{jos@iac.es}
        \and
        Departamento de Astrof\'\i sica, Universidad
        de La Laguna, Tenerife, Spain
	\and
	Kwasan and Hida Observatories, 
	Kyoto University, Yamashina, 
	Kyoto 607-8471, Japan\\
	\email{ichimoto@kwasan.kyoto-u.ac.jp}
	}

   \date{Received~~~~~/ Accepted~~~~~}
  
 
  \abstract
	{The topology of penumbral magnetic fields 
	is poorly known. 
	The satellite {{\em Hinode}} has recently revealed 
	penumbral structures with 
	a magnetic polarity opposite to the main sunspot 
	polarity. They may be a direct confirmation 
	of magnetic field lines and mass flows returning
	to the solar interior throughout the penumbra, a 
	configuration previously inferred from interpretation 
	of observed Stokes profile asymmetries. 
}
	{To point out the relationship between the reverse polarity 
	features found by {\em Hinode}, and the model 
	Micro-Structured Magnetic Atmospheres (MISMAs)
	proposed for sunspots.
}
   {Synthesis and modeling the sunspot Stokes profiles.}
   {Existing model MISMAs produce strongly redshifted
	reverse polarity structures as found by {\em Hinode}. 
	Ad hoc model MISMAs  
	also explain 
	the asymmetric Stokes 
	profiles observed by {\em Hinode}. 
	The same modeling may be consistent with magnetograms of 
	dark cored penumbral filaments if the dark cores are 
	associated with the reverse polarity. Such hypothetical 
	relationship will show up only in the far red wings 
	of the spectral lines.
}
   {The reverse polarity patches 
        may result from 
 	aligned magnetic field lines and mass flows that
	bend over and return to the solar interior throughout the penumbra.
}
\keywords{ 
	Sun: magnetic fields --
	Sun: photosphere --
	sunspots
	}

\authorrunning{S\'anchez~Almeida \& 
	Ichimoto}
\titlerunning{Origin of reverse polarities
	in penumbrae}
\maketitle
\section{Rationale}\label{introduction}

Sunspots have always been benchmarks 
to test our understanding of magneto 
convection. It is known for long 
that magnetic forces impede the
free plasma motions, thus reducing the
efficiency of convection \citep{bie41,cow53}. 
However, 
convection occurs in sunspots despite the
strength of the magnetic field, and the 
high conductivity of the  photospheric
plasma. The problem arises as to what is the 
mode of convection, i.e., how
magnetic fields and plasma flows
adjust one another to allow
transporting the energy that balances 
the radiation losses. 
This long-lasting problem is far from 
been settled, and it is particularly
severe in the penumbrae of sunspots with 
predominantly horizontal magnetic fields 
and mass flows
\citep[for recent reviews see, e.g.,][]{sol03,tho04,sch08,san09,tho09}. 
From an observational point of view, the problem
lies in the small physical
scales at which the convective transport 
is organized. Even 
with the best spatial resolutions achieved 
at present, we cannot follow the rise, cooling,
and subsequent submergence of plasma blobs.
The topology of the magnetic field lines and 
flows must be inferred indirectly. 
The present paper is devoted to analyze
and interpret a recent observation  
that may be central to constrain the 
topology of the magnetic fields in penumbrae. 
 
\citet{ich07} report the presence 
of a strongly redshifted magnetic component in the
penumbrae of sunspots with a polarity opposite to the
main sunspot polarity. This component shows up throughout 
the penumbra, a property used to argue 
that the redshift must be due to vertical velocities. 
The observations were carried out with the stallelite
{\em Hinode} \citep[][]{kos07},
which yields a spatial resolution of 0\farcs 32  
at the working wavelength 
\citep[$\simeq$\,6302\,\AA;][]{tsu08}.
\,\citeauthor{ich07} 
finding seems to be at variance with the magnetograms 
taken by \citet{lan05,lan07} with the Swedish Solar Tower
\citep[SST,][]{sch02}, which do not show magnetic fields of reverse polarity 
in penumbrae. 
This poses a serious problem since SST has 
twice {\em Hinode} spatial resolution and, therefore,
it should be simpler for SST to resolve and 
detect mixed polarities. SST magnetograms only
reveal a decrease of 
the magnetograph signals coinciding with the 
dark cores, i.e., the dark lanes outlined 
by bright filaments discovered by \citet{sch02}. 
This work shows how {\em Hinode} and SST observations 
can be naturally understood within the two 
component semi-empirical model penumbra derived 
by \citet[][hearafter SA05]{san04b},
provided that the dark cores are associated 
with the reverse polarity.

SA05 works out the model 
MIcro-Structured Magnetic Atmospheres (MISMAs\footnote{
The acronym was coined by \citet{san96} 
to describe magnetic atmospheres having optically-thin
substructure, which naturally produce asymmetric 
spectral lines.})
required to quantitatively 
reproduce the asymmetries of the 
Stokes profiles\footnote{As usual, the Stokes parameters 
are used to characterize
the polarization; $I$ for the intensity,
$Q$ and $U$ for the two independent types
of linear polarization, and $V$ for the
circular polarization. The 
Stokes profiles are 
graphs of $I$, $Q$, $U$ and $V$ versus 
wavelength for a particular spectral line.
They follow well defined symmetries when
the atmosphere has constant magnetic field 
and  velocity 
\citep[see, e.g.,][]{lan92}.} 
observed in a large sunspot. 
Other inversion techniques have succeed in 
reproducing the observed line shapes 
\citep[e.g.,][]{san92b,wes01a,mat03}, 
but there is something unique to the MISMA inversion, 
namely, the model demands two opposite 
polarities. This unexpected result
has been often criticized as unreal 
\citep[e.g.,][]{lan06,bel09}, however, 
it is the ingredient that 
naturally explains \hinode\ reversals.
The MISMA model sunspot includes two 
magnetic components.
The major component 
contains most of the mass of each resolution element, 
and it has the polarity of the sunspot.  
It is generally combined with a minor component
of opposite polarity and having large velocities.
In typical 1\arcsec-resolution observations,  the outcoming
light is systematically dominated by the major component, 
and the resulting Stokes profiles have rather regular 
shapes. An exception occurs in the so-called 
{\em apparent neutral line}, where 
the Stokes~$V$ profiles show a characteristic shape with 
three or more lobes termed {\em cross-over effect} 
\citep[see][and references therein]{san92b}. 
At the neutral line the mean magnetic 
field vector is perpendicular to the 
line-of-sight, and 
the contribution  of the major component almost 
disappears 
in Stokes~$V$
due to projection effects. 
The cancellation of the two components is expected to be
less effective when improving the spatial resolution,
leading to
the appearance of cross-over profiles. 
Actually, {\em Hinode} often finds 
cross-over Stokes $V$ profiles, and they show
up precisely at the location
of the reverse polarities
\citep[Fig.~5 in][and also \S~\ref{hinode}]{ich07}.
The observed cross-over profiles have two polarities:
the main sunspot polarity close to the line 
center, and the reverse polarity at the far red wing.
Since the
reverse polarity patches detected in penumbrae 
by {\em Hinode} 
produce cross-over profiles, they
seem to correspond to structures where the
polarity is not well defined, with positive
and negative polarities coexisting in each pixel.

The paper is structured as follows: 
\S~\ref{model} shows how the model MISMAs 
from \paperiii\ qualitatively
reproduce both {\em Hinode} and SST observations. 
We work out a simple model penumbral filament to show that it 
grasps the essential features of the observed ones. The
same agreement is found  when the model MISMAs are inferred
by fitting
actual 
{\em Hinode} Stokes profiles (\S~\ref{hinode}).  
The implications in the context of the 
penumbral magnetic field topology and the Evershed 
effect are discussed in \S~\ref{discussion}, where
we also put forward a specific test that could
confirm or falsify our explanation.
Empirical and theoretical 
difficulties for the dark cores to be
associated with the reverse polarities
are also discussed in \S~\ref{discussion}.

%
%
\section{Model MISMA for penumbral filaments with
	dark core}\label{model}

\begin{figure*}
\centering
\includegraphics[width=0.8\textwidth]{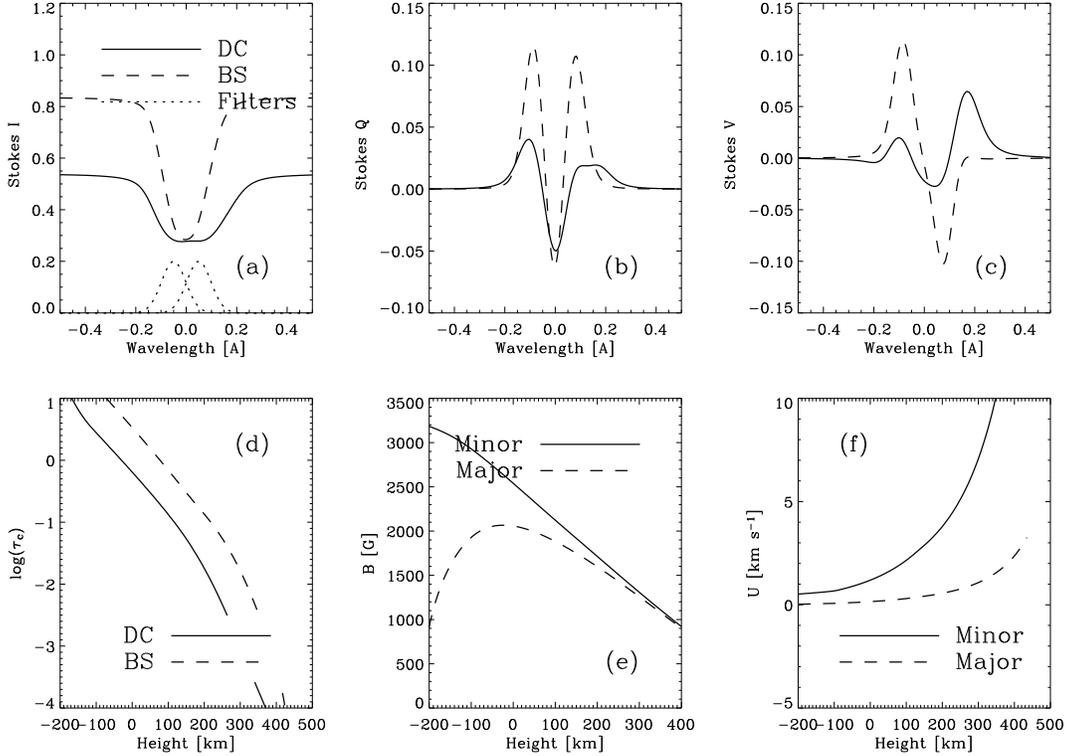}
\caption{
(a) Stokes~$I$ profiles in one of the representative
model MISMAs in \paperiii , which has been
slightly modified to represent a dark core 
(the solid line), and its bright sides (the dashed line).
They are normalized to the quiet Sun continuum intensity.
(b) Stokes~$Q$ profiles. 
(c) Stokes~$V$ profiles.
(d) Continuum optical depth $\tau_c$ vs height in the
	atmosphere for the dark core and the bright
sides, as indicated in the inset.
(e) Magnetic field strength vs height for the two
magnetic components of
the model MISMAs. They are identical for 
the dark core and  the bright sides.
(f) Velocities along the magnetic field lines for
the two magnetic components of the model MISMAs.
They are identical for 
the dark core and the bright sides. 
}
\label{almeida-hinode0}
\end{figure*}

As we describe in the
introductory section, the model MISMAs often
require two magnetic components 
with opposite polarities to
reproduce the observed Stokes profiles. 
The major component has the polarity 
of the sunspot, and it is  combined with a 
minor component of opposite polarity and having 
large velocities. 
The outcoming
light is dominated by the major component, so that
the reverse polarity seldom produces an
obvious signal in the spatially
integrated Stokes profiles.
Within this scenario, improving 
the spatial resolution would reduce the
spatial smearing, allowing extreme cases 
to show up.
In order to mimic the effect of  improving
spatial resolution,
several randomly chosen model MISMAs in \paperiii\ 
were modified by  increasing the fraction of 
atmosphere occupied by the minor component.
Now the minor
\footnote{
Here and throughout, {\em minor} and {\em major}
refer to the two components in the  model sunspot by SA05. 
When applied to the components in the model atmospheres 
worked out in the paper, it only implies that their magnetic 
and velocity properties are similar to the minor and 
major components in SA05.
} 
component  shows up in Stokes~$V$. 
The behavior described next is common
to all the models, but we only examine in detail
the example given in 
Fig.~\ref{almeida-hinode0}.  
The resulting Stokes~$I$, $Q$ and $V$
profiles 
of {Fe}~{\sc i}~$\lambda$6302.5~\AA\ are represented
as solid lines in Figs.~\ref{almeida-hinode0}a,
\ref{almeida-hinode0}b and \ref{almeida-hinode0}c, 
respectively. They correspond 
to a point in the 
limb-side penumbra of a sunspot at $\mu=0.95$ (18$^\circ$
heliocentric angle). 
Note how Stokes~$I$ is redshifted and
deformed, and how  Stokes~$V$ shows the cross-over
effect. 
Consequently, the improvement of spatial resolution
with respect to traditional earth-based spectro-polarimetric
observations naturally explains the abundance 
of cross-over Stokes~$V$ profiles found by {\em Hinode}. 
Figures~\ref{almeida-hinode0}a,  \ref{almeida-hinode0}b,
and \ref{almeida-hinode0}c
also show the case where the major component
dominates (the dashed line). The strong
asymmetries have disappeared, rendering 
Stokes~$V$ with reasonably 
antisymmetric  shape and the 
sign of the dominant
polarity. 
Recall that the two sets of Stokes profiles
in Figs.~\ref{almeida-hinode0}a,
\ref{almeida-hinode0}b, and 
\ref{almeida-hinode0}c
(the solid lines and the dashed lines)
have been produced with exactly
the same magnetic field vectors
and mass flows (shown in 
Figs.~\ref{almeida-hinode0}e and  \ref{almeida-hinode0}f).
The atmospheres differ because of the relative importance 
of major and minor components, and because of a global 
scaling factor in the temperature stratification.
One of them is 
some 80\% cooler than the other one.
The coolest 
renders 
asymmetric profiles with low continuum
intensity, suitable to 
mimic dark features (see the Stokes~$I$ continua 
in Fig.~\ref{almeida-hinode0}a).

\begin{figure}
\centering
\includegraphics[width=0.55\textwidth]{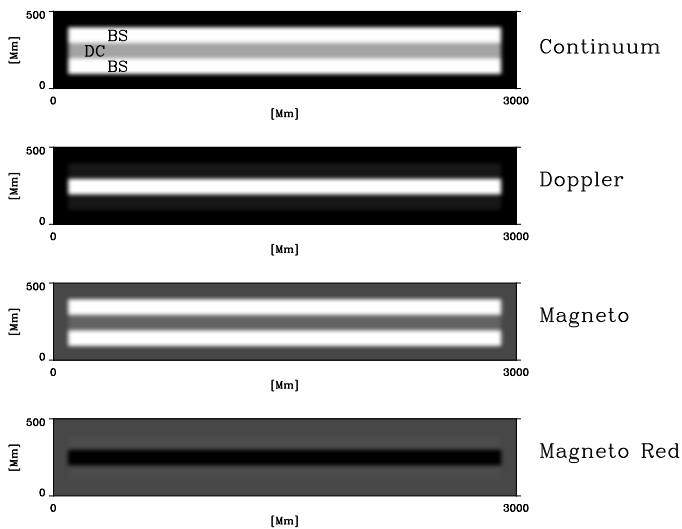}
\caption{
Schematic modeling of SST observations
of penumbral filaments by \citet{lan05,lan07}.
A dark core (DC) surrounded by two
bright sides (BS) 
is located in the limb-side 
penumbra of a sunspot at $\mu=0.95$
(i.e., 18$^\circ$ heliocentric angle).
The three top images show a
continuum image, a dopplergram,
and a magnetogram, as labelled. 
The convention is such that both the 
sunspot main polarity
and a redshift produce positive signals.
The dark background in all
images has been included for reference, and
it corresponds to signal equals zero. 
The fourth image ({\tt Magneto Red}) 
corresponds to a magnetogram in the far red wing 
of Fe~{\sc i}~$\lambda$6302.5~\AA , and it reveals a 
dark core with a polarity opposite
to the sunspot main polarity.
The continuum image and the dopplergram
have been scaled from zero (black) to 
maximum (white). The scaling of the two
magnetograms is the same, so that their 
signals can be compared directly.
}
\label{almeida-hinode1}
\end{figure}

Understanding {\em Hinode} observations in terms 
of MISMAs also explains the lack of reverse polarity
in SST magnetograms.  Stokes~$V$ in 
reverse polarity regions shows cross-over
effect (Fig.~5 in \citealt{ich07}, and
the solid line in Fig.~\ref{almeida-hinode0}c),
i.e.,
it presents two polarities depending on the 
sampled wavelength. 
It has the main sunspot polarity
near line center, whereas the polarity 
is reversed in the far red wing.
SST magnetograms are taken at line center 
($\pm 50$\ m\AA ), which explains why the 
reverse polarity does not show up.
A significant reduction of
the Stokes~$V$ signal occurs, though. Such reduction
naturally explains the observed
weakening of magnetic signals in 
dark cores 
\citep[][ and \S~\ref{introduction}]{lan05,lan07}, 
provided that
the dark cores are associated with an 
enhancement of the opposite polarity,
i.e., if the dark cores produce cross-over profiles. 
In order to illustrate the argument, we have 
constructed images, magnetograms, and dopplergrams 
of a (na\"\i ve) model dark-cored filament. 
It is formed by a uniform 100 km wide dark strip,
representing the dark core,  
bounded by two bright strips of the same width,
representing the bright sides.  
The Stokes profiles of the dark core have been 
taken as the solid lines in Figs.~\ref{almeida-hinode0}a 
and \ref{almeida-hinode0}c, whereas the bright sides are modelled
as the dashed lines in the same figures.
The color filters employed 
by \citet{lan05,lan07}
are approximated by Gaussian functions 
of 80\,m\AA\ FWHM, and shifted 
$\pm 50$\,m\AA\ from the line center 
(see the dotted lines in Fig.~\ref{almeida-hinode0}a).
The magnetogram signals are computed from the
profiles as  
\begin{equation}
M=-{{\Delta\lambda}\over{|\Delta\lambda|}}
	{{{\int V(\lambda) f(\lambda-\Delta\lambda)\,d\lambda}}\over{
	\int I(\lambda) f(\lambda-\Delta\lambda)\,d\lambda}}, 
\end{equation}
with the wavelength $\lambda$ referred
to the central wavelength of the line,
$f(\lambda)$ the transmission curve of the
filter centered at $\lambda=0$, and $\Delta\lambda=-50$\,m\AA . 
Similarly, the Doppler signals are given by
\begin{equation}
D={{\Delta\lambda}\over{|\Delta\lambda|}}
{{\int I(\lambda)\, [f(\lambda+\Delta\lambda)-f(\lambda-\Delta\lambda)]\,d\lambda}
	\over {\int I(\lambda)\, [f(\lambda+\Delta\lambda)+
	f(\lambda-\Delta\lambda)]\,d\lambda}},
\end{equation}
but here we employ the Stokes~$I$ profile of the non-magnetic
line used by 
\citeauthor{lan07}~(\citeyear{lan07}; 
i.e., {Fe}~{\sc i}~$\lambda$5576~\AA).
When $\Delta\lambda < 0$,
the signs of $M$ and $D$ ensure 
$M >0$ for the main polarity of the sunspot, 
and $D> 0$ for redshifted profiles.
The continuum intensity 
has been taken as $I$ at -0.4\,{\rm \AA} from the
line center. The continuum image of this model filament
is shown in Fig.~\ref{almeida-hinode1}, with the dark core 
and the bright sides marked as DC and BS, respectively.
The dopplergram and the magnetogram are also included in the same
figure. The dark background in all images indicates 
the level corresponding to no signal.
In agreement with \citeauthor{lan07} observations,
the filament shows redshifts ($D > 0$), which are 
enhanced in the dark core. 
In agreement with \citeauthor{lan07},
the filament shows the main polarity of the sunspot
($M > 0$), with the signal
strongly reduced in the dark core.
Figure~\ref{almeida-hinode1}, bottom, 
includes the 
magnetogram to be observed at the far red wing
($\Delta\lambda=200$\,m\AA). The dark core now shows
the reverse polarity ($M< 0$), whereas the bright 
sides still maintain the main polarity with an 
extremely weak signal. This specific
prediction of the modeling is amenable
for direct observational test (see \S~\ref{discussion}).

Two additional remarks on our modeling
are in order. First, the magnetogram signal
in the dark core is much weaker than  in the
bright sides, despite the fact that the (average)
magnetic field strength is larger in the core
(see Fig.~\ref{almeida-hinode0}e, keeping in mind that
the minor component dominates).
Second, the model dark core is depressed
in height
with respect to the bright sides. 
Figure~\ref{almeida-hinode0}d shows the continuum 
optical depth $\tau_c$ as a function of the 
height in the atmosphere. When the two  atmospheres 
are in lateral pressure balance, 
the layer $\tau_c=1$ of the
dark core is shifted by some 100~km downward
with respect to the same layer in the 
bright sides. 
The depression of the observed
layers in the dark core is  produced by two 
conspiring effects; 
the decrease of density associated with the increase 
of magnetic pressure \citep[e.g.,][]{spr76}, 
and the decrease of opacity associated
with the reduction of temperature \citep[e.g.,][]{sti91}.
We mention the association of our dark cores with enhanced 
field strength and with geometrical depression because these
properties contrast with some popular models of penumbral
magneto-convection \citep[e.g.][]{sch06,rem08}.
Note, however, that the association between field strength, 
brightness and geometric height is far from being 
established.
Not all models predict dark features coinciding with weaker
field. The siphon flow model of the Evershed effect has 
enhanced field strengths in the downflowing leg 
\citep[e.g.,][]{sch02d}. The plasma has already cool down 
when reaching this footpoint and, so, one expects 
downflows associated with stronger fields and colder 
plasmas. 
As for the elevation of the dark cores, there is a solid
observational result disfavoring it. 
\citet{sch04} found that the limb-side penumbra and 
the center-side penumbra are darker than the rest. They 
interpret this observation as an effect of the depression of 
the dark penumbral filaments, which are obscured
by the bright ones when observed sideways.
Obviously,  this is an average result, but it strongly 
suggests that if dark cores are common, then they must be 
depressed with respect to the bright sides. 
We are showing here how the reduced magnetograph 
signals observed in dark cores can be produced 
even if their field strength is enhanced. 
%
%
\section{Reproducing {\em Hinode} Stokes profiles}\label{hinode}

We have gone a step further, and the 
exercise in the previous section has been repeated
using model MISMAs
derived directly from {\em Hinode}/SP data. 
We use a small
set of seven Stokes profiles selected so that they
represent  extreme cases among
the whole range  of redshifted and blueshifted profiles.
They were observed  in a simple, positive-polarity 
sunspot (NOAA10944) when the target was almost at
the solar disk center (heliocentric angle 1\fdg 1), so that the
line-of-sight direction and the vertical direction 
coincide to most purposes.
The observation is described in \citet{ich08}, 
and we refer to this work for images, a
logbook, and further details. 
Our data correspond 
to those taken at 18:25 UT 
on February 28th, 2007.
Normal scan maps were obtained with the 
Spectro-Polarimeter (SP) of the Solar
Optical Telescope 
\citep[SOT;][]{tsu08,sue08} 
aboard Hinode \citep{kos07}.
The SP took full Stokes profiles of  
\linea\  and \lineb\ with 0.1\% photometric accuracy, 
and a spatial sampling of  0\farcs 16.

\begin{figure*}
\centering
\includegraphics[width=.8\textwidth]{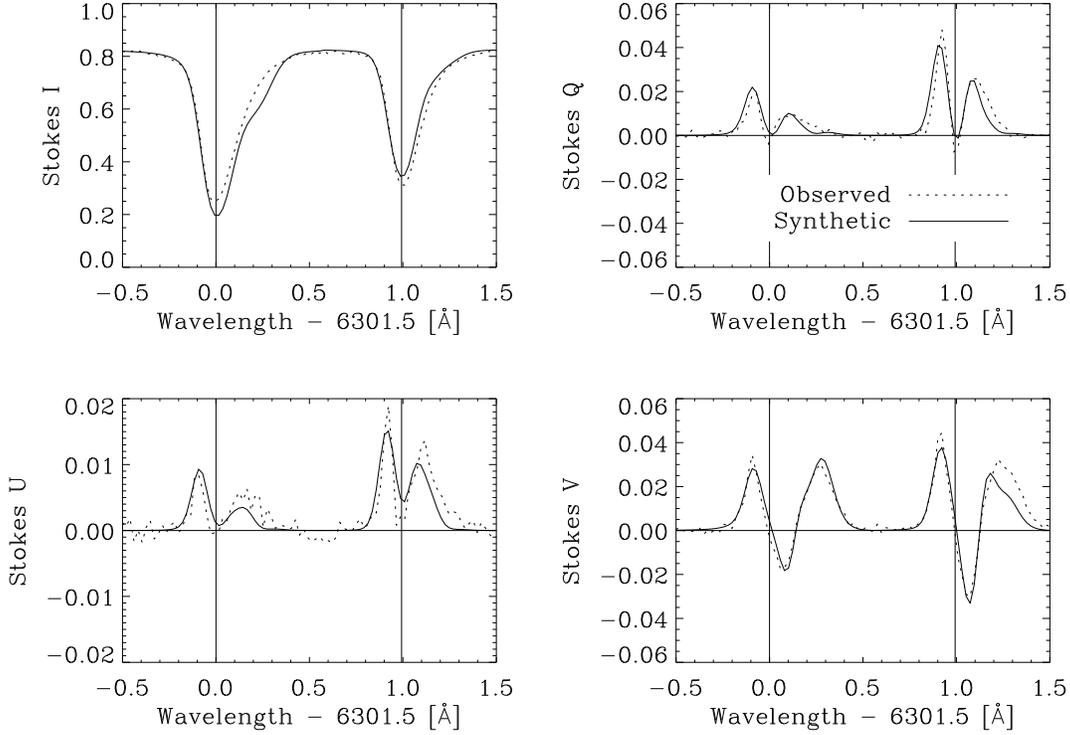}
\caption{
Set of Stokes profiles of \linea\ and \lineb\
observed by {\em Hinode} in one of the reverse polarity regions (the dotted 
lines).
(The ordinate axis 
labels identify the Stokes parameter.) 
Note how Stokes~$V$
shows the cross-over effect (i.e., three lobes rather than two).
The solid lines correspond to a MISMA inversion of this
set of profiles, and it renders the model atmosphere shown 
in Fig.~\ref{r1_model}.
Wavelengths are referred to the laboratory wavelength
of \linea . The vertical solid lines indicate the
laboratory wavelengths of \linea\ and \lineb .
}
\label{r1_fit}
\end{figure*}

\begin{figure*}
\centering
\includegraphics[width=.8\textwidth]{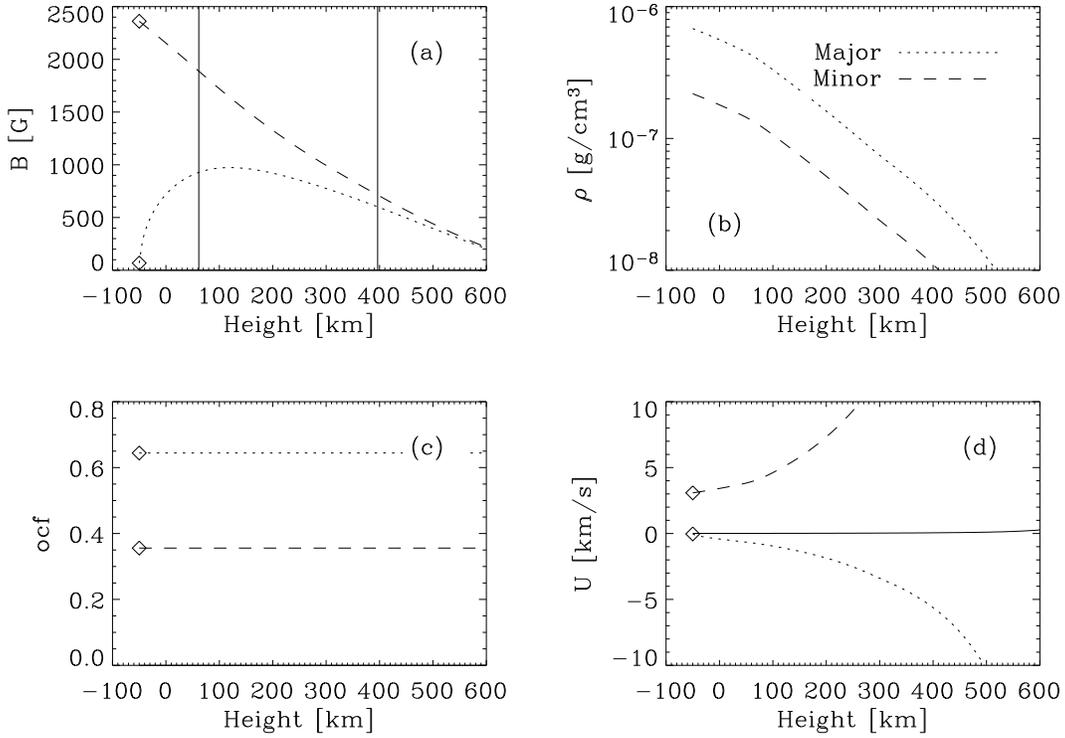}
\caption{
Model MISMA for one of the typical 
cross-over profiles.
It has been derived from inversion of the profiles 
shown in Fig.~\ref{r1_fit}, the dotted lines.
(a) Stratification of magnetic field strength.
As the inset in (b) indicates, the minor and the
major components can be identified by the type 
of line.
(b) Stratification of density. (c) Fraction of the
atmosphere occupied by the two components.
(d) Stratification of velocity along magnetic field lines.
Note that in order to get the Doppler shifts,
the velocity $U$ has to be corrected for the 
inclination of the magnetic field. 
As the minor and major components have opposite 
polarities, both yield redshifts.
(The magnetic field inclinations 
of the major 
and minor components are  
74$^\circ$\ and 144$^\circ$, respectively,
so that the major component has positive polarity
whereas the minor component has negative polarity.
)
The symbols correspond to the quantities
used as free parameters during fitting,
which set the full stratification of the atmosphere
via MHD constraints \citep{san97b}.
}
\label{r1_model}
\end{figure*}

\begin{figure*}
\centering
\includegraphics[width=.8\textwidth]{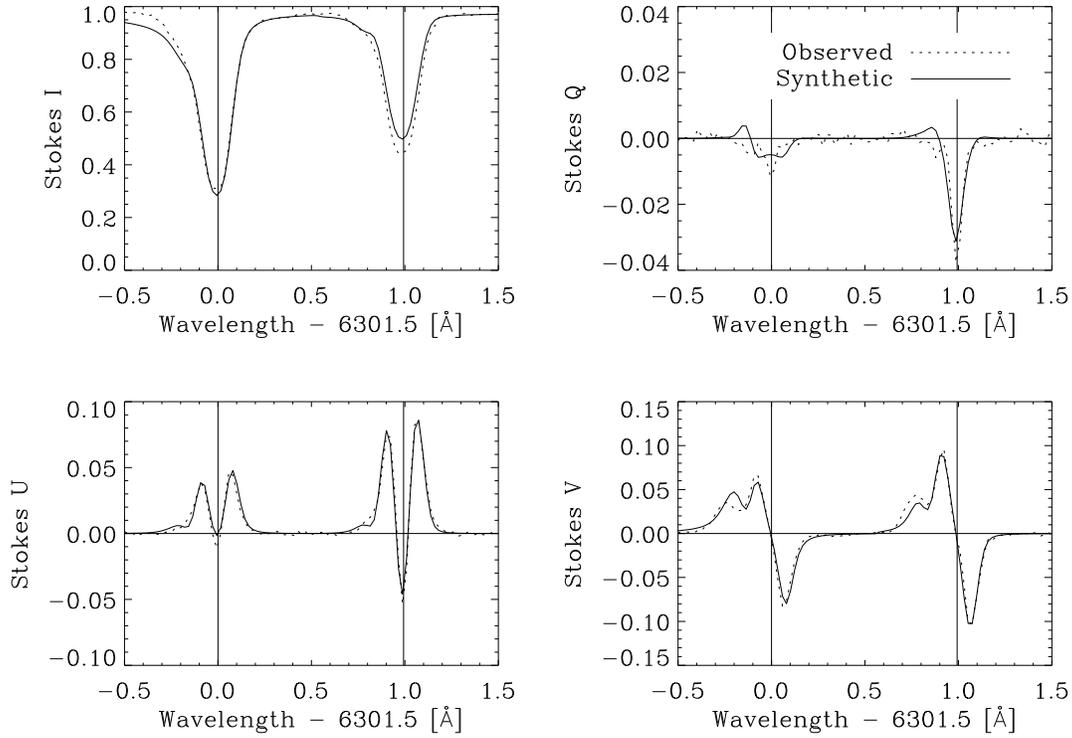}
\caption{
Same as Fig.~\ref{r1_fit} but for one of the
strongly blueshifted regions. In this case the
solid lines correspond to the model atmosphere in
Fig.~\ref{b4_model}.
}
\label{b4_fit}
\end{figure*}

\begin{figure*}
\centering
\includegraphics[width=.8\textwidth]{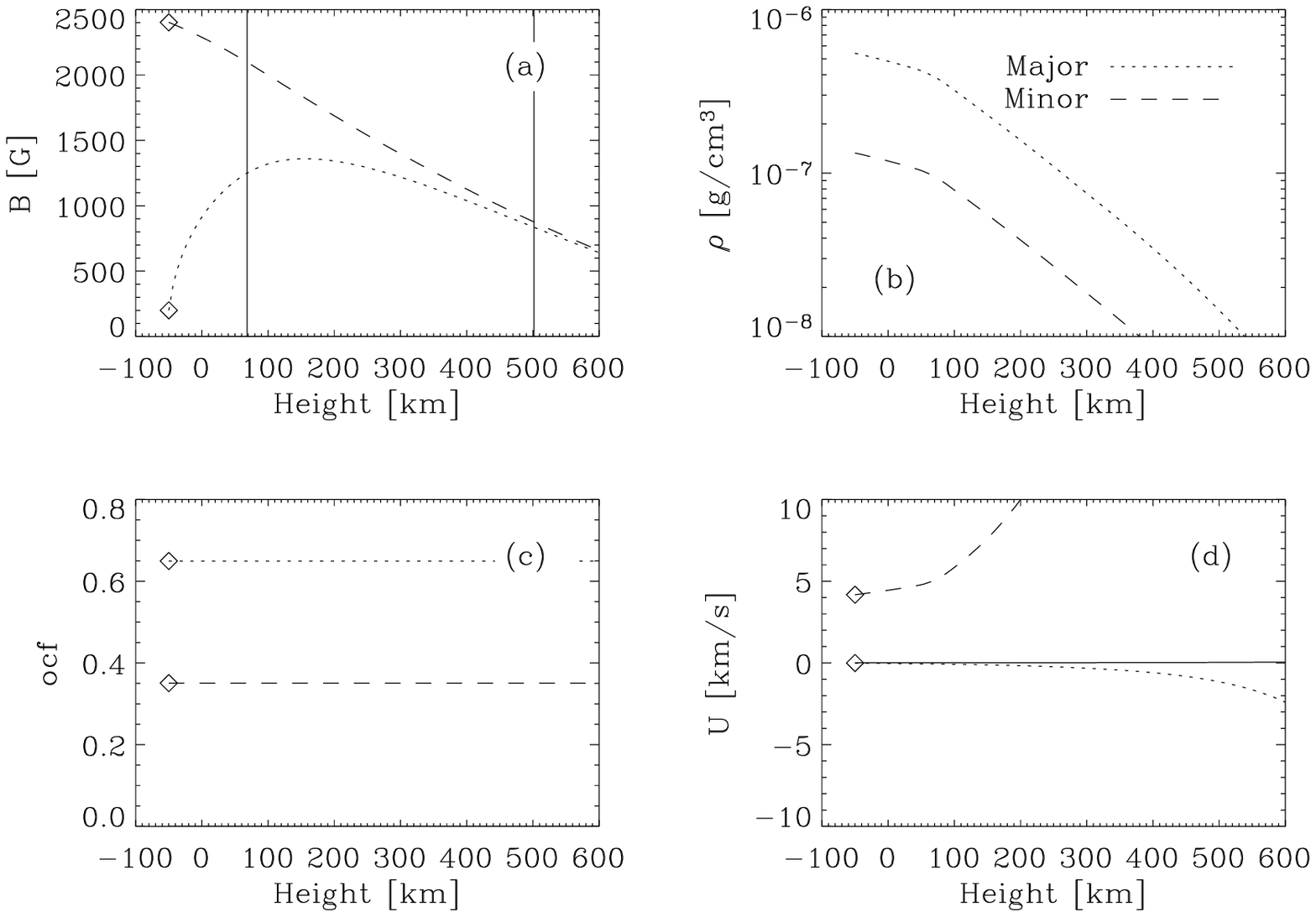}
\caption{
Model MISMA reproducing one of the typical blueshifted 
profiles (the dotted lines in Fig.~\ref{b4_fit}).
See caption of Fig.~\ref{r1_model} for 
a description of the various plots.
The inclinations 
of the major 
and minor components are  
66$^\circ$\ and 39$^\circ$, respectively,
so that both components have positive polarity.
After correcting $U$ for the magnetic field 
inclination,
the velocity of the minor component 
along a vertical line-of-sight corresponds
to blueshifts. The change 
with respect to Fig.~\ref{r1_model} is due
to the flip of the minor component polarity, 
which is negative in Fig.~\ref{r1_model} and 
positive here.
}
\label{b4_model}
\end{figure*}

The MISMA inversion procedure described in \citet{san97b} 
provides fair fits in all cases. 
Two examples are shown in Figs.~\ref{r1_fit} 
and \ref{b4_fit}. The dotted lines in Fig.~\ref{r1_fit} 
correspond to one representative reverse polarity site.
The fit, shown as solid lines, yields the model
atmosphere represented in Fig.~\ref{r1_model}.  
The inversions were carried out as described
in \paperiii , and we refer to that paper for 
details. The only significant difference 
was the setting
up of the absolute wavelength scale, which we zeroed from
the average intensity profile in a quiet Sun 
region far from the sunspot. The wavelength of the
core of \lineb\ is assumed to correspond to a
global velocity equals to the convective blueshift
of the line measured by \citet{dra81}.
Figures~\ref{b4_fit} and \ref{b4_model} are
similar to Figs.~\ref{r1_fit} and \ref{r1_model},
except that they represent a point with clear blueshift.
The model atmospheres are similar,
except for the important detail that the 
minor component does not have reverse polarity 
in the case of these strongly blueshifted profiles.
The four panels in Figs.~\ref{r1_model} 
and \ref{b4_model} represent the stratification 
with height in the atmosphere
of (a) magnetic field strength, (b) density,
(c) fraction of atmosphere occupied by each
component, and (d) velocity 
along magnetic field lines. The minor
component occupies a significant fraction of the 
atmosphere (some 40\% in the examples
in the figures), and it has low density and
high field strength. Field strengths and
densities are similar to those found
in \paperiii,
however, the fraction of atmosphere 
occupied by the minor component is 
significantly higher (almost twice the 
typical 20\% in \paperiii ).
This is to be expected since we have selected for 
inversion pixels with particularly large asymmetries,
where the contribution of the minor component
must exceed the average to cause a significant
impact on the Stokes profiles.


\begin{figure}
\includegraphics[width=0.55\textwidth]{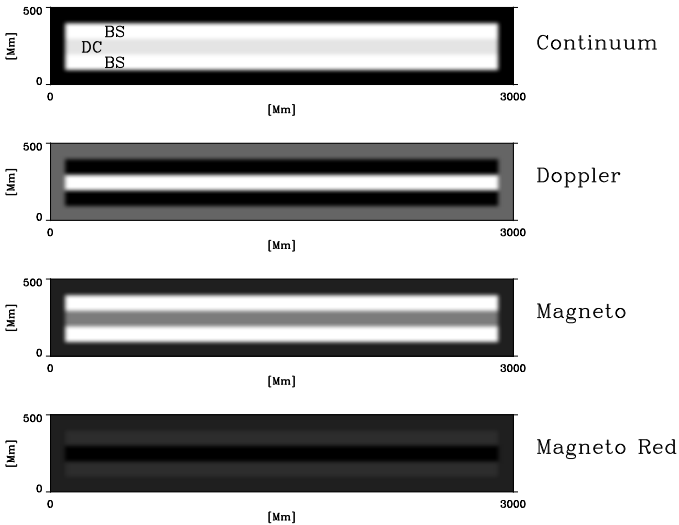}
\caption{
Set of synthetic images, dopplergrams and magnetograms
equivalent to Fig~\ref{almeida-hinode1}, but
using the model MISMAs directly derived from {\em Hinode}
spectra to represent the dark core and its bright sides.
The main features remain as in Fig~\ref{almeida-hinode1}. 
}
\label{hinode3}
\end{figure}

We have repeated the exercise leading to the synthetic
magnetograms and dopplergrams in Fig.~\ref{almeida-hinode1},
but using the model MISMAs in Figs.~\ref{r1_model} and \ref{b4_model}.
Specifically, we use the profiles in the blueshifted region to 
represent the bright sides, 
and the redshifted profiles for the 
dark core.
The result is shown in Fig.~\ref{hinode3}. The main features
of  Fig.~\ref{almeida-hinode1} remain: (1) the bright
filament has a dark core, (2) the line-center
magnetogram has a weakening coinciding with the dark core,
(3) Bright sides are blueshifted with respect to the dark cores,
and (4) the far red wing magnetogram
shows opposite polarity coinciding
with the dark core. Note that features 1--3 are in agreement 
with SST observations.

A clarification may be appropriate.
The association between the Stokes
profiles in Fig.~\ref{r1_fit} and dark cores, 
and the profiles in Fig.~\ref{b4_fit} and 
bright sides is a mere working hypothesis.  
\hinode\ spectra barely resolve 
bright sides and dark cores
and, therefore, the used profiles 
do not correspond to identifiable 
bright sides and dark cores. 
We have selected them because 
they illustrate the properties to be expected 
for bright sides and dark cores 
according to the modeling in 
\S~\ref{model}. Even with limitted 
resolution, just by chance, 
some pixels may have enhanced contribution of 
bright sides and dark cores.

%
%
\section{Discusion}\label{discussion}

The model MISMAs by SA05 predict and produce
strongly redshifted reverse polarity 
structures 
similar to those
found by {\em Hinode}
in penumbrae  (\S~\ref{introduction}).
In addition to pointing out this agreement
(\S~\ref{model}),
we have applied the kind of modelling employed
by SA05 to quantitatively reproduce some representative 
very asymmetric Stokes profiles observed by 
{\em Hinode} (\S~\ref{hinode}).
In order to fit the
Stokes~$V$ profiles with three lobes observed 
in reverse polarity regions 
(cross-over profiles), the model MISMAs 
have two
components of opposite polarity in each 
resolution element. The minor component holds the 
reverse polarity, and it always carries strong magnetic-field-aligned 
flows. 
High spatial resolution SST magnetograph observations of 
penumbra do not show reverse polarities. They just 
indicate a 
weakening of the Stokes~$V$ signal coinciding
with the dark cores \citep[][]{lan05,lan07}.
The presence of reverse polarities and 
the absence of their signals in SST 
magnetograms can be explained if the dark cores are 
associated with reverse polarity Stokes~$V$ profiles. 
Such association has not been revealed so far because the 
existing magnetograms were taken at line core, 
whereas the reverse polarity only shows up at the
far red wing of the spectral lines. 
By tuning the bandpass of SST magnetograms to 
the appropriate wavelength, this specific prediction 
of our modeling can be tested 
observationally\footnote{Even if our prediction 
turns out to be incorrect, 
the disagreement between {\em Hinode} and SST magnetograms
is a serious problem that urges solution.}. 
The association between  cross-over Stokes~$V$ 
profiles and dark cores is still a mere conjecture. However, 
we would like to mention an independent Hinode/SP observation
which also suggests such association.
\citet[][]{bel07} find and discuss the case of 
a limb-side penumbra dark core which clearly shows
cross-over Stokes~$V$ profiles  (see their Fig.~2). 
As we argue in the introduction, Hinode/SP spatial 
resolution does not suffice to properly resolve 
dark cores, but \citeauthor{bel07} observation is 
encouraging.
It indicates that the dark cores are associated
with several magnetic field inclinations 
in the resolution element.

The magnetic fields that we use in \S~\ref{hinode}
to reproduce the observed 
asymmetries are rather horizontal.
However, the flows along field lines are so intense 
(in excess of 10 km~s$^{-1}$ in the minor component; 
see the botton right
 panels in Figs.~\ref{r1_model} 
and \ref{b4_model}) that
the vertical component of the velocities are
of the order of a few~km\,s$^{-1}$.  
Order of magnitude estimates show that
1~km\,s$^{-1}$
suffices to explain the transport of energy 
by convection in penumbra
\citep[e.g., ][]{spr87,stei98,san09}.
For the transport to be effective, 
vertical velocities of such magnitude should 
be present everywhere. It is still unclear whether 
such large velocities
are common enough to be responsible for the 
required convective transport, and such possibility 
has to be studied in detail.

{\em Hinode} observations show the 
reversals to prefer the outer penumbra, whereas  
asymmetric blueshifted profiles (like those in
Fig.~\ref{b4_fit})
cluster toward the inner penumbra \citep{ich07}.
One might think that this fact compromises our 
interpretation, because dark cores tend to appear in the 
inner penumbra. However, beware of oversimple 
interpretations of {\em Hinode} spectra. 
{\em Hinode}/SP does not resolve individual dark cores and 
bright sides, but spatially integrate them. 
If the bright sides dominate, then the
resulting Stokes~$V$ profiles would show properties 
of the bright sides, even if the pixel contains a dark core. Actually, 
blueshifted profiles seem to be associated 
with bright continuum features, supporting that
the bright sides dominate in these points. As one 
moves toward the outer penumbra, the magnetic field of the bright sides
becomes more horizontal, reducing the Stokes~$V$ signals, 
and  allowing the dark core polarization to show up. In 
agreement with this conjecture, 
the reverse polarity patches seem to coincide with dark lanes
\citep{ich07}.

Insufficient resolution can also be 
invoked to explain the different morphological 
appearance of the \hinode\ polarity
reversals and the SST dark cores. Dark cores are 
elongated features, 
but the locations of downflows described by \citet{ich07} 
are much more point-like features. Cross-over Stokes~$V$ 
signals critically depend on the orientation of the magnetic 
fields with respect to the line-of-sight. Small
modifications of the magnetic field geometry
would make the reverse polarity unobservable,
overwhelmed by the Stokes~$V$ signals of 
the main sunspot polarity. The reverse polarity
shows up only when a number of conditions are met, 
but this need to satisfy several delicate tradeoffs
makes the presence of identifiable reverse polarities
rare, therefore, they tend to be spatially scattered, 
discontinuous, and so point-like. On the contrary, 
these delicate balances do not affect the intensity and, 
therefore, unpolarized light SST images tend to 
show more continuous structures with the filamentary
appearance characteristic of penumbrae.

A final important comment is in order.
Stokes~$V$ profiles like those in Figs.~\ref{r1_fit} and 
\ref{b4_fit} have net circular polarization (NCP), i.e.,
the wavelength integral of the Stokes~$V$ profile differs
from zero. NCP can {\em only} be produced by
variation of the magnetic field and velocity
{\em along} the line-of-sight,  
a well known fact from the early works by 
\citet{ill75} and \citet{aue78}  trying to explain the
broad-band circular polarization in 
sunspots \citep[see, e.g.,][ and references therein]{san92b}. 
This implies that strongly asymmetric Stokes~$V$ profiles
showing NCP will be always present in sunspots, 
no matter the spatial resolution of the observation 
{\em across} the line-of-sight.
Even if we improve the resolution of our telescope
to infinity, we will never be able to separate the
penumbrae into pixels where the magnetic field is uniform. 
In other words, resolving the fine scale structure of 
the penumbral magnetic 
field is not (only) a question of improving the
spatial resolution, but it requires understanding 
 line asymmetries.
Whether this understanding requires MISMAs or can
be accomplished with a smoother magnetic field
distribution is still a matter of debate
(e.g., SA05, Sect.~5; \citeauthor{lan06}~\citeyear{lan06}).

\begin{acknowledgements} 

{\em Hinode} is a Japanese mission developed and launched by ISAS/JAXA, 
with NAOJ as domestic partner and NASA and STFC (UK) as 
international partners. It is operated by these agencies 
in co-operation with ESA and NSC (Norway).
The work has partly been funded by the Spanish Ministry of Science
and Technology, project AYA2007-66502, as well as by
the EC SOLAIRE Network (MTRN-CT-2006-035484).
This work was partly carried out at the 
NAOJ Hinode Science Center, which is 
supported by the Grant-in-Aid for Creative Scientific Research, 
The Basic Study of SpaceWeather Prediction from MEXT, 
Japan (Head Investigator: K. Shibata), 
generous donations from Sun Microsystems, 
and NAOJ internal funding.
\end{acknowledgements}


\begin{thebibliography}{36}
\expandafter\ifx\csname natexlab\endcsname\relax\def\natexlab#1{#1}\fi

\bibitem[{{Auer} \& {Heasley}(1978)}]{aue78}
{Auer}, L.~H. \& {Heasley}, J.~N. 1978, \aap, 64, 67

\bibitem[{{Bellot Rubio}(2009)}]{bel09}
{Bellot Rubio}, L.~R. 2009, in Magnetic Coupling between the Interior and the
  Atmosphere of the Sun, ed. S.~S. {Hassan} \& R.~J. {Rutten}, Astrophysics and
  Space Science Proceedings (Berlin: Springer-Verlag), in press,
  arXiv:0903.3619 [astro-ph]

\bibitem[{{Bellot Rubio} {et~al.}(2007){Bellot Rubio}, {Tsuneta}, {Ichimoto},
  {Katsukawa}, {Lites}, {Nagata}, {Shimizu}, {Shine}, {Suematsu}, {Tarbell},
  {Title}, \& {del Toro Iniesta}}]{bel07}
{Bellot Rubio}, L.~R., {Tsuneta}, S., {Ichimoto}, K., {et~al.} 2007, \apjl,
  668, L91

\bibitem[{{Biermann}(1941)}]{bie41}
{Biermann}, L. 1941, Vierteljahrsschr. Astr. Gesellsch., 76, 194

\bibitem[{{Cowling}(1953)}]{cow53}
{Cowling}, T.~G. 1953, {Solar Electrodynamics}, ed. G.~P. {Kuiper} (Chicago:
  The University of Chicago Press), 532

\bibitem[{{Dravins} {et~al.}(1981){Dravins}, {Lindegren}, \&
  {Nordlund}}]{dra81}
{Dravins}, D., {Lindegren}, L., \& {Nordlund}, A.~. 1981, \aap, 96, 345

\bibitem[{{Ichimoto} {et~al.}(2007){Ichimoto}, {Shine}, {Lites}, {Kubo},
  {Shimizu}, {Suematsu}, {Tsuneta}, {Katsukawa}, {Tarbell}, {Title}, {Nagata},
  {Yokoyama}, \& {Shimojo}}]{ich07}
{Ichimoto}, K., {Shine}, R.~A., {Lites}, B., {et~al.} 2007, \pasj, 59, 593

\bibitem[{{Ichimoto} {et~al.}(2008){Ichimoto}, {Tsuneta}, {Suematsu},
  {Katsukawa}, {Shimizu}, {Lites}, {Kubo}, {Tarbell}, {Shine}, {Title}, \&
  {Nagata}}]{ich08}
{Ichimoto}, K., {Tsuneta}, S., {Suematsu}, Y., {et~al.} 2008, \aap, 481, L9

\bibitem[{{Illing} {et~al.}(1975){Illing}, {Landman}, \& {Mickey}}]{ill75}
{Illing}, R.~M.~E., {Landman}, D.~A., \& {Mickey}, D.~L. 1975, \aap, 41, 183

\bibitem[{{Kosugi} {et~al.}(2007){Kosugi}, {Matsuzaki}, {Sakao}, {Shimizu},
  {Sone}, {Tachikawa}, {Hashimoto}, {Minesugi}, {Ohnishi}, {Yamada}, {Tsuneta},
  {Hara}, {Ichimoto}, {Suematsu}, {Shimojo}, {Watanabe}, {Shimada}, {Davis},
  {Hill}, {Owens}, {Title}, {Culhane}, {Harra}, {Doschek}, \& {Golub}}]{kos07}
{Kosugi}, T., {Matsuzaki}, K., {Sakao}, T., {et~al.} 2007, \solphys, 243, 3

\bibitem[{{Landi Degl'Innocenti}(1992)}]{lan92}
{Landi Degl'Innocenti}, E. 1992, in Solar Observations: Techniques and
  Interpretation, ed. F.~{S\'anchez}, M.~{Collados}, \& M.~{V\'azquez}
  (Cambridge: Cambridge University Press), 71

\bibitem[{{Langhans}(2006)}]{lan06}
{Langhans}, K. 2006, in ASP Conf. Ser., Vol. 358, Solar Polarization 4, ed.
  R.~{Casini} \& B.~W. {Lites} (San Francisco: ASP), 3

\bibitem[{{Langhans} {et~al.}(2005){Langhans}, {Scharmer}, {Kiselman},
  {L\"ofdahl}, \& {Berger}}]{lan05}
{Langhans}, K., {Scharmer}, G., {Kiselman}, D., {L\"ofdahl}, M., \& {Berger},
  T.~E. 2005, \aap, 436, 1087

\bibitem[{{Langhans} {et~al.}(2007){Langhans}, {Scharmer}, {Kiselman}, \&
  {L{\"o}fdahl}}]{lan07}
{Langhans}, K., {Scharmer}, G.~B., {Kiselman}, D., \& {L{\"o}fdahl}, M.~G.
  2007, \aap, 464, 763

\bibitem[{{Mathew} {et~al.}(2003){Mathew}, {Lagg}, {Solanki}, {Collados},
  {Borrero}, {Berdyugina}, {Krupp}, {Woch}, \& {Frutiger}}]{mat03}
{Mathew}, S.~K., {Lagg}, A., {Solanki}, S.~K., {et~al.} 2003, \aap, 410, 695

\bibitem[{{Rempel} {et~al.}(2009){Rempel}, {Sch\"ussler}, \&
  {Kn\"olker}}]{rem08}
{Rempel}, M., {Sch\"ussler}, M., \& {Kn\"olker}, M. 2009, \apj, 691, 640

\bibitem[{{S\'anchez Almeida}(1997)}]{san97b}
{S\'anchez Almeida}, J. 1997, \apj, 491, 993

\bibitem[{{S\'anchez Almeida}(2005)}]{san04b}
{S\'anchez Almeida}, J. 2005, \apj, 622, 1292

\bibitem[{{S{\'a}nchez Almeida}(2009)}]{san09}
{S{\'a}nchez Almeida}, J. 2009, in Magnetic Coupling between the Interior and
  the Atmosphere of the Sun, ed. S.~S. {Hassan} \& R.~J. {Rutten}, Astrophysics
  and Space Science Proceedings (Berlin: Springer-Verlag), in press,
  arXiv:0902.4532 [astro-ph]

\bibitem[{{S\'anchez Almeida} {et~al.}(1996){S\'anchez Almeida}, {Landi
  Degl'Innocenti}, {Mart\'\i nez Pillet}, \& {Lites}}]{san96}
{S\'anchez Almeida}, J., {Landi Degl'Innocenti}, E., {Mart\'\i nez Pillet}, V.,
  \& {Lites}, B.~W. 1996, \apj, 466, 537

\bibitem[{{S\'anchez Almeida} \& {Lites}(1992)}]{san92b}
{S\'anchez Almeida}, J. \& {Lites}, B.~W. 1992, \apj, 398, 359

\bibitem[{{Scharmer} {et~al.}(2002){Scharmer}, {Gudiksen}, {Kiselman}, {L{\"
  o}fdahl}, \& {Rouppe van der Voort}}]{sch02}
{Scharmer}, G.~B., {Gudiksen}, B.~V., {Kiselman}, D., {L{\" o}fdahl}, M.~G., \&
  {Rouppe van der Voort}, L.~H.~M. 2002, \nat, 420, 151

\bibitem[{{Scharmer} \& {Spruit}(2006)}]{sch06}
{Scharmer}, G.~B. \& {Spruit}, H.~C. 2006, \aap, 460, 605

\bibitem[{{Schlichenmaier}(2002)}]{sch02d}
{Schlichenmaier}, R. 2002, Astron. Nachr., 323, 303

\bibitem[{{Schlichenmaier}(2009)}]{sch08}
{Schlichenmaier}, R. 2009, \ssr, 213

\bibitem[{{Schmidt} \& {Fritz}(2004)}]{sch04}
{Schmidt}, W. \& {Fritz}, G. 2004, \aap, 421, 735

\bibitem[{{Solanki}(2003)}]{sol03}
{Solanki}, S.~K. 2003, \aapr, 11, 153

\bibitem[{{Spruit}(1976)}]{spr76}
{Spruit}, H.~C. 1976, \solphys, 50, 269

\bibitem[{{Spruit}(1987)}]{spr87}
{Spruit}, H.~C. 1987, in The Role of Fine-Scale Magnetic Fields on the
  Structure of the Solar Atmosphere, ed. E.-H. {Schr\"o ter}, M.~{V\'aquez}, \&
  A.~A. {Wyller} (Cambridge: Cambridge University Press), 199

\bibitem[{{Stein} \& {Nordlund}(1998)}]{stei98}
{Stein}, R. F.~I. \& {Nordlund}, {\AA}. 1998, \apj, 499, 914

\bibitem[{{Stix}(1991)}]{sti91}
{Stix}, M. 1991, The Sun (Berlin: Springer-Verlag)

\bibitem[{{Suematsu} {et~al.}(2008){Suematsu}, {Tsuneta}, {Ichimoto},
  {Shimizu}, {Otsubo}, {Katsukawa}, {Nakagiri}, {Noguchi}, {Tamura}, {Kato},
  {Hara}, {Kubo}, {Mikami}, {Saito}, {Matsushita}, {Kawaguchi}, {Nakaoji},
  {Nagae}, {Shimada}, {Takeyama}, \& {Yamamuro}}]{sue08}
{Suematsu}, Y., {Tsuneta}, S., {Ichimoto}, K., {et~al.} 2008, \solphys, 249,
  197

\bibitem[{{Thomas}(2009)}]{tho09}
{Thomas}, J.~H. 2009, in Magnetic Coupling between the Interior and the
  Atmosphere of the Sun, ed. S.~S. {Hassan} \& R.~J. {Rutten}, Astrophysics and
  Space Science Proceedings (Berlin: Springer-Verlag), in press,
  arXiv:0903.4106 [astro-ph]

\bibitem[{{Thomas} \& {Weiss}(2004)}]{tho04}
{Thomas}, J.~H. \& {Weiss}, N.~O. 2004, \araa, 42, 517

\bibitem[{{Tsuneta} {et~al.}(2008){Tsuneta}, {Ichimoto}, {Katsukawa}, {Nagata},
  {Otsubo}, {Shimizu}, {Suematsu}, {Nakagiri}, {Noguchi}, {Tarbell}, {Title},
  {Shine}, {Rosenberg}, {Hoffmann}, {Jurcevich}, {Kushner}, {Levay}, {Lites},
  {Elmore}, {Matsushita}, {Kawaguchi}, {Saito}, {Mikami}, {Hill}, \&
  {Owens}}]{tsu08}
{Tsuneta}, S., {Ichimoto}, K., {Katsukawa}, Y., {et~al.} 2008, \solphys, 249,
  167

\bibitem[{{Westendorp Plaza} {et~al.}(2001){Westendorp Plaza}, {del Toro
  Iniesta}, {Ruiz Cobo}, {Mart\'\i nez Pillet}, {Lites}, \&
  {Skumanich}}]{wes01a}
{Westendorp Plaza}, C., {del Toro Iniesta}, J.~C., {Ruiz Cobo}, B., {et~al.}
  2001, \apj, 547, 1130

\end{thebibliography}

\end{document}